# Particle-wave dichotomy in quantum Monte Carlo: unlocking the quantum correlations

Ivan P. Christov

Physics Department, Sofia University, 1164 Sofia, Bulgaria

Email: ivan.christov@phys.uni-sofia.bg

**Abstract**

Here, a dichotomy of particles and waves is employed in a quantum Monte Carlo calculation of interacting electrons. Through the creation and propagation of concurrent stochastic ensembles of walkers in physical space and in Hilbert space one can correctly predict the ground state and the real-time evolution of a single electron interacting with larger quantum system. It is shown that such walker ensembles can be constructed straightforwardly through a stochastic sampling (windowing) applied to the mean-field approximation. Our calculations reveal that the ground state and the real-time evolution of the probability distributions and the decoherence due to the Coulomb interaction in presence of strong ultrashort laser pulse can be accounted for correctly by calculating the density matrix of the electron, without referencing to the quantum many-body state of the whole system.

## 1. Introduction

The recent experimental progress of laser technologies triggered substantial research in light-matter interaction in ultrafast time scale with theoretical efforts directed towards better understanding of the intimate mechanisms which control the motion of electrons in atoms, molecules and condensed matter [1]. Significant part of that research focuses on *ab initio* methods, where in the ideal case the direct (numerical) solution of the time-dependent Schrödinger equation would provide not only the ground state but also the real-time evolution of the many-body wave-function which next would allow all quantities of interest to be calculated. Unfortunately, the wave function is a physical object only for an isolated particle, while in the many-body case it is defined in configuration space which introduces both formal and computational difficulties in the quantum mechanics framework in that the solution of the many-body quantum problem by a classical computer requires effort which scales exponentially with the system size. Therefore substantial progress in both methodology and numerical algorithms would be needed to deliver accurate and numerically efficient methods for real time evolution of many-body quantum systems.

While obvious for the many-body Schrödinger equation the exponential scaling may concern also the particle based methods (e.g. quantum Monte Carlo) and, in general, any method pursuing exact solution [2,3]. It is expected therefore that much efforts have been devoted to reformulate quantum mechanics in terms of physical space (one-body) wave functions which are easier to calculate under different conditions as required by the experiment. The most prominent of those are the Hartree-Fock and the density functional approximations. The reduction of the many-body quantum problem to lower dimensionality comes, however, at the price of losing essential quantum properties such as the quantum correlations due to the particle interactions which are simply neglected in Hartree-Fock and introduced semi-empirically in the density functional. This leads, for example, to inaccurate estimations of the ground state and/or binding energies of some atoms and molecules. One class of improved accuracy many-body methods involves multiple configurations (e.g. in [4]) which can be considered to be generalization of Hartree-Fock where the quantum correlations can be accounted for. However these methods are significantly less numerically efficient for realistic systems which may involve multiple electrons. The situation is somewhat improved with the help of quantum Monte Carlo (QMC) methods (e.g. the diffusion QMC, [5,6]) where variationally optimized but essentially artificial many-body trial functions of Slater-Jastrow-type are used to guide the diffusing Monte Carlo walkers to their most energetically favorable positions in the presence of interaction potentials. In this way very accurate estimates for the ground and excited state energies have been obtained, but the real-time evolution still remains beyond reach. Besides that, for convergence of the results for fermions, additional constrains have to be imposed close to the nodes of the ground state wave-function (fixed node approximation), which are aimed to deal with the infamous "sign problem". Somewhat complementary to the diffusion QMC is the auxiliary field QMC (AFQMC) where the random walk occurs in the state (or wave-function) space where an imaginary-time propagator projects the trial state to the ground state [7]. At the heart of AFQMC lies the continual Fourier transform (Hubbard-Stratonovich transformation) which recasts the system of interacting particles to a system of non-interacting particles moving in stochastic fields. However, for repulsive interactions (e.g. for those between each two electrons) the stochastic potentials turn to be complex quantities, which is counter-intuitive for the ground state and causes a "phase problem" analogous to the "sign problem" above. The essentially stochastic nature of AFQMC also prevents its use for real-time propagation. In an attempt to address in a numerically efficient way some of the existing problems in the treatment of interacting quantum particles for both imaginary- and real-time propagation a new quantum Monte Carlo method was introduced recently [8,9]. The method was named time dependent quantum Monte Carlo (TDQMC) and it involves concurrently ensembles of particles and wave-functions, such that these evolve self-consistently in physical space-time. The symmetric way in which particles and waves are treated in TDQMC is called here particle-wave dichotomy. The method is well suited for calculation of both ground state properties and real-time evolution of quantum averages and coherences for many-body quantum systems.

This paper provides a systematic derivation of the working equations of the TDQMC method and presents numerical results for the ground state and the time evolution of a simple quantum system. The results for one-dimensional helium atom in external field are compared with those from the exact numerical solution of the many-body Schrödinger equation.

## 2. Unlocking the quantum correlations from the mean-field approximation

### 2.1. Time-dependent quantum Monte Carlo

We start here with the many-body Schrödinger equation for *N*-electron atom in an external field:

$$i\hbar\frac{\partial}{\partial t}\Psi(\mathbf{R},t)=-\frac{\hbar^2}{2m}\nabla^2\Psi(\mathbf{R},t)+V(\mathbf{R},t)\Psi(\mathbf{R},t) \tag{1}$$

where $\mathbf{R}=(\mathbf{r}_1,\mathbf{r}_2,...,\mathbf{r}_N)$ and $\nabla=(\nabla_1,\nabla_2,...,\nabla_N)$. The potential $V(\mathbf{R},t)$ in Eq. (1) is a sum of electron-nuclear, electron-electron, and external potentials:

$$V(\mathbf{r}_1,...,\mathbf{r}_N,t)=V_{e-n}(\mathbf{r}_1,...,\mathbf{r}_N)+V_{e-e}(\mathbf{r}_1,...,\mathbf{r}_N)+V_{ext}(\mathbf{r}_1,...,\mathbf{r}_N,t)$$

$$=\sum_{k=1}^{N}V_{e-n}(\mathbf{r}_k)+\sum_{k>l}^{N}V_{e-e}(\mathbf{r}_k-\mathbf{r}_l)+V_{ext}(\mathbf{r}_1,...,\mathbf{r}_N,t)\,. \tag{2}$$

From here, we use as a starting point the mean-field approximation which can be derived either variationally or by substituting a simple representation of the many-body wave function expressed as a product of one-body wave functions:

$$\Psi(\mathbf{r}_1,\mathbf{r}_2,...,\mathbf{r}_N,t)=\prod_{i=1}^{N}\varphi_i\,(\mathbf{r}_i,t) \tag{3}$$

in Eq.(1) and next integrating over the spatial variables except $\mathbf{r}_i$:

$$i\hbar\frac{\partial}{\partial t}\varphi_i(\mathbf{r}_i,t)=\left[-\frac{\hbar^2}{2m}\nabla_i^2+V_{e-n}(\mathbf{r}_i)+V_{ext}(\mathbf{r}_i,t)+V_{e-e}^{Hartree}(\mathbf{r}_i,t)\right]\varphi_i(\mathbf{r}_i,t)\,, \tag{4}$$

where:

$$V_{e-e}^{Hartree}\left(\mathbf{r}_i,t\right)=\sum_{j\neq i}^{N}\frac{1}{\int d\mathbf{r}_j\left|\varphi_j(\mathbf{r}_j,t)\right|^2}\int d\mathbf{r}_j\left|\varphi_j(\mathbf{r}_j,t)\right|^2 V_{e-e}(\mathbf{r}_i-\mathbf{r}_j)\qquad , \tag{5}$$

is the time dependent Hartree potential and we have neglected some independent on $\mathbf{r}_i$ phase terms in the RHS of Eq.(4). The fact that we assign a single wave function $\varphi_i(\mathbf{r}_i,t)$ to the *i*-th electron implies that we presume that electron be in a pure state. It is clear however that going beyond the mean-field approximation, the interaction of each electron with the rest of the electrons should, in principle, lead to a mixture of states due to the correlated electron motion, which remains hidden in the mean-field because the Hartree potential depends on the entire distribution $\left|\varphi_j(\mathbf{r}_j,t)\right|^2$ for the *j*-th electron in Eq.(5). One simple intuitive way to unlock the electron-electron correlations from the mean-field approximation would be to introduce randomness in the local environment experienced by each electron. Accordingly, it is assumed in TDQMC that portions of the *i*-th electron cloud may experience the Coulomb potential due to random parts of the *j*-th electron cloud and vice versa. One way to accomplish this is to introduce a stochastic windowing of $\left|\varphi_j(\mathbf{r}_j,t)\right|^2$ in Eq. (5) by a kernel function $K\left[\mathbf{r}_j,\mathbf{r}_j^k(t),\sigma_j^k\right]$ with a width $\sigma_j^k$, centered at a point $\mathbf{r}_j^k(t)$, such that the product $\left|\varphi_j(\mathbf{r}_j,t)\right|^2 K\left[\mathbf{r}_j,\mathbf{r}_j^k(t),\sigma_j^k\right]$ is different for each separate $\mathbf{r}_j^k(t)$. For example, for Gaussian kernel we have:

$$K\left[\mathbf{r}_j,\mathbf{r}_j^k(t),\sigma_j^k\right]=\exp\left(-\frac{\left|\mathbf{r}_j-\mathbf{r}_j^k(t)\right|^2}{2\sigma_j^k\left(\mathbf{r}_j^k,t\right)^2}\right) \tag{6}$$

Then, the windowing kernel transforms the probability density under the integrals in Eq.(5) into a mixture of densities, according to the rule:

$$\left|\varphi_j(\mathbf{r}_j,t)\right|^2\Rightarrow\left|\varphi_j^k(\mathbf{r}_j,t)\right|^2\Rightarrow\left|\varphi_j^k(\mathbf{r}_j,t)\right|^2 K\left[\mathbf{r}_j,\mathbf{r}_j^k(t),\sigma_j^k\right], \tag{7}$$

where starting from the mean-field probability density for the j-the electron we first introduce a set of probability densities for an ensemble of wave functions assigned to that electron in order to account for the quantum mixture which arises due to the quantum correlations. The second step in the chain of Eq.(7) defines the "windowing" where $\left|\varphi_j^k(\mathbf{r}_j,t)\right|^2 K\left[\mathbf{r}_j,\mathbf{r}_j^k(t),\sigma_j^k\right]$ is the "window" is space centered at $\mathbf{r}_j^k(t)$ and weighted by $\left|\varphi_j^k(\mathbf{r}_j,t)\right|^2$. Then, substituting Eq.(7) into Eq.(5) introduces a new (effective) potential:

$$V_{e-e}^{eff}\left(\mathbf{r}_i,t\right)=\sum_{j\neq i}^{N}\frac{1}{\int d\mathbf{r}_j\left|\varphi_j^k(\mathbf{r}_j,t)\right|^2 K\left[\mathbf{r}_j,\mathbf{r}_j^k(t),\sigma_j^k\right]}\int d\mathbf{r}_j\left|\varphi_j^k(\mathbf{r}_j,t)\right|^2 K\left[\mathbf{r}_j,\mathbf{r}_j^k(t),\sigma_j^k\right]V_{e-e}\left(\mathbf{r}_i-\mathbf{r}_j\right), \tag{8}$$

which is different for each $k$ and therefore transforms the mean-field equation, Eq.(4), into a set of equations for the different replicas $\varphi_i^k(\mathbf{r}_i,t)$ of the original one-body wave function $\varphi_i(\mathbf{r}_i,t)$:

$$i\hbar\frac{\partial}{\partial t}\varphi_i^k(\mathbf{r}_i,t)=\left[-\frac{\hbar^2}{2m}\nabla_i^2+V_{e-n}(\mathbf{r}_i)+V_{ext}(\mathbf{r}_i,t)+V_{e-e}^{eff}(\mathbf{r}_i,t)\right]\varphi_i^k(\mathbf{r}_i,t) \tag{9}$$

The next key assumption in TDQMC is that, in fact, the trajectories $\mathbf{r}_j^k(t)$ may be represented by a set of $M$ Monte Carlo walkers in physical space, each of which samples its own probability density given by $\left|\varphi_j^k(\mathbf{r}_j,t)\right|^2$ where $\varphi_j^k(\mathbf{r}_j,t)$ represents a concurrent walker in Hilbert space. This allows us to transform the integrals in Eq.(8) into sums according to the Monte Carlo integration rule, thus allowing to express the effective potential as a Monte Carlo convolution:

$$V_{e-e}^{eff}\left(\mathbf{r}_i,t\right)=\sum_{j\neq i}^{N}\frac{1}{Z_j^k}\sum_{l}^{M}V_{e-e}\left[\mathbf{r}_i-\mathbf{r}_j^l(t)\right]K\left[\mathbf{r}_j^l(t),\mathbf{r}_j^k(t),\sigma_j^k\right], \tag{10}$$

where:

$$Z_j^k = \sum_l^M K\left[\mathbf{r}_j^l(t), \mathbf{r}_j^k(t), \sigma_j^k\right]. \tag{11}$$

Finally, the TDQMC equations for the wave functions, Eqs.(9)-(11), need to be complemented by equations which connect the motion for the Monte Carlo walkers in physical space with those in Hilbert space, where the most natural choice are the de Broglie-Bohm guiding equations (see e.g. in [10]):

$$\mathbf{v}_i^k = \frac{d\mathbf{r}_i^k}{dt} = \frac{\hbar}{m}\mathrm{Im}\left[\frac{1}{\varphi_i^k(\mathbf{r}_i,t)}\nabla_i\varphi_i^k(\mathbf{r}_i,t)\right]_{\mathbf{r}_i=\mathbf{r}_i^k(t)}, \tag{12}$$

where $k$=1,…,$M$ denote the Monte Carlo walkers which describe the $i$-th quantum particle; $i$=1,…,$N$. Having in mind the historical origin of Eq.(12) the walkers in Hilbert space will be called, henceforth, the “guide waves”. Note that similar guide equations have been used for decades in diffusion QMC for imaginary time propagation where the velocity of the walkers was named “quantum force” [5]. In this way, equations (9) through (12) present a method to evolve concurrent ensembles of particles and guide waves in a self-consistent manner in space and time.

Since the windowing parameter $\sigma_j^k$ in Eq.(10) determines the amount of walkers belonging to the $j$-th electron which contribute to the electron-electron potential experienced by the $i$-th electron, the parameter $\sigma_j^k$ is named in TDQMC the nonlocal quantum correlation length. Clearly, in the limiting case where $\sigma_j^k \to 0$, the set of TDQMC equations (9) is transformed into a set of linear Schrödinger equations with a classical pair-wise potential $V_{e-e}^{eff}(\mathbf{r}_i,t) = V_{e-e}\left[\mathbf{r}_i, \mathbf{r}_j^k(t)\right]$, which basically overestimates the electron-electron repulsion. The opposite case where $\sigma_j^k \to \infty$ returns us to the mean-field approximation which underestimates the electron-electron repulsion. While the results are not expected to be sensitive to the exact shape of the kernel in Eq. (10), for e.g. bell-shaped kernels, it is physically justified to specify the parameter $\sigma_j^k$ to be proportional to some global statistical parameter which characterizes the coordinate uncertainty for each separate quantum particle, such as the kernel density estimation bandwidth (or simply the standard deviation $\sigma_j(t)$) of the corresponding Monte Carlo sample, and to be inversely proportional to the average distance between the walkers for the different electrons. In case of localized distributions for most of the time (e.g. in an atom at ground state), one can write:

$$\sigma_j^k\left(\mathbf{r}_j^k, t\right) = \alpha_j . \sigma_j(t), \tag{13}$$

where $\alpha_j$ is a variational parameter of the order of unity which is determined by minimizing the ground state energy of the system under consideration. Once specified, the parameter $\alpha_j$ remains constant during the real-time evolution while $\sigma_j^k$ changes automatically with the spread of the wave function. For practical implementation, it should be noted that a Monte Carlo sampling occurs during the preparation of the ground state of the system through a Markovian diffusion process, consistent with the kinetic part of Eq.(9) (see also [5]):

$$d\mathbf{r}_i^k = \mathbf{v}_i^k dt + \boldsymbol{\eta}(t)\sqrt{\frac{\hbar}{m} dt}\ , \tag{14}$$

where $\mathbf{v}_i^k$ is the walker velocity calculated from Eq.(12) and $\boldsymbol{\eta}(t)$ is a vector random variable with zero mean and time-dependent variance which tends to zero towards steady state. Clearly, for imaginary time propagation the velocity $\mathbf{v}_i^k$ in Eq.(14) vanishes from Eq. (12). Note that the random motion of the walkers gives rise to randomness of the kernel windowing function of Eq. (10) which may thus be considered as a stochastic auxiliary field to participate in the one-body Schrodinger equation for the guide waves (Eq. (9)). However, unlike in AFQMC the resulting potential in Eq.(10) stays real for all times and it is stabilized towards the end of the ground state propagation stage which is of crucial importance for the stability of the whole method. Also, an importance sampling that involves branching (birth and death of couples of Monte Carlo particles and waves) was found to greatly improve the convergence and the accuracy of the TDQMC calculation during the preparation stage, similarly to in other quantum Monte Carlo methods [5]. Once steady state is established the walkers are next guided by the corresponding guide waves (see Eq.(12)) for real-time propagation. This contrasts other approaches which use stochastic wave-functions, e.g. in the quantum jump method [11,12], where significant stochastic component is introduced also during the real time propagation of the quantum system, which limits the propagation time to the so called "useful time" [13].

### 2.2. Statistics of particles and waves

We have seen so far that the quantum particle-wave dichotomy can be reformulated for interacting particles where instead of stating that each physical particle is guided by a guide wave we can say that each physical particle is represented by a set of fictitious particles (walkers) whose distribution in physical space reproduces its probability density, and each of those walkers is guided by its own wave function (guide wave), as drawn schematically in Fig.1. Then, as indicated in Fig.1, in the mean-field approximation all walkers are guided by the same wave function and the electron is in a pure state. In other words, together with an ensemble of walkers in physical space, for each physical particle there exists a joined statistical ensemble of guide waves which differ from each other if that particle is in a mixed state. This, on the other hand, warrants the introduction of appropriate density matrices and correlation functions based on these guide waves. The definition of the density matrix in our case is simplified by the fact that the classical probability for occurrence of a given wave function (guide wave) from the mixture is imposed by the distribution of the Monte Carlo walkers in space and therefore we may construct the density matrix for each separate electron directly, e.g. by using a simple mean over the guide waves [14]:

$$\rho_i\left(\mathbf{r}_i,\mathbf{r}'_i,t\right)=\frac{1}{M}\sum_{k=1}^{M}\varphi_i^{k*}(\mathbf{r}_i,t)\varphi_i^{k}(\mathbf{r}'_i,t) \quad , \tag{15}$$

without having to solve the von Neumann equation or the Lindblad equation. This is a significant progress since the standard calculation of the density matrix for a subsystem through a reduction of the density matrix of the whole system would require a lot more operations. Also, the above definition complies with the properties of a density matrix and it provides correct values for the operator averages in coordinate representation. From a statistical viewpoint, the density matrix describes important properties of the ensemble of guide waves considered as random variables in Hilbert space, namely, it serves as the variance-covariance matrix where different definitions of a variance of an operator can be introduced (see e.g. [12], Ch.3). It is clear, therefore, that the one-body density matrix of Eq.(15) may provide an instrumental approach to access probability densities and coherences for parts of a complex quantum system without having to access the many-body quantum state. For example, averages such as the dipole moment or the atomic ionization of an electron exposed to an external electric field can be calculated using either the walkers in physical space or those in Hilbert space (the guide waves) provided by the TDQMC methodology, as demonstrated in the next section.

Besides for the calculation of densities (distributions) the walker trajectories can be conveniently used to calculate various many-body effects without having to calculate multi-dimensional integrals. For example, starting from the two-electron density:

$$\rho_2\left(\mathbf{r}_1,\mathbf{r}_2,t\right)=\int\left|\Psi(\mathbf{R},t)\right|^2 d\mathbf{r}_3...d\mathbf{r}_N \ , \tag{16}$$

one can determine the probability of two electrons being a given distance apart [15]:

$$P(\mathbf{u},t)=\int \rho_2\left(\mathbf{r}_1,\mathbf{r}_2,t\right)\delta\left[\left(\mathbf{r}_1-\mathbf{r}_2\right)-\mathbf{u}\right]d\mathbf{r}_1 d\mathbf{r}_2 , \tag{17}$$

which, from Eq.(3), gives after a simple integration (for spherical symmetry):

$$P(u,t)\propto\sum_{k=1}^{M}\mathrm{K}_k\left[\frac{\left|r_{12}^k\left(t\right)-u\right|}{\sigma_{12}}\right], \tag{18}$$

where $r_{12}^k(t) = \left|\mathbf{r}_1^k(t) - \mathbf{r}_2^k(t)\right|$; $i$=1…$M$. In this way the pair density function for the electrons can be reduced to simple kernel density estimation with kernel $K_k$ and bandwidth $\sigma_{12}$ over the ensemble of the distances between the corresponding Monte Carlo walkers.

## 3. Numerical results

Before we present the results of the numerical implementation of TDQMC and discuss the results a few remarks are needed. As already mentioned Eq.(14) accomplishes the diffusion of the walkers in physical space where each walker samples its own guide wave and where the walkers velocity tend to zero at the ground state. It is important that the time dependence of the stochastic variable $\boldsymbol{\eta}(t)$ is not too abrupt because the walkers would not manage to thermalize to the correct distribution. It was found that linear dependence $\boldsymbol{\eta}(t) = (\tau - t)\boldsymbol{\eta}_0$ is appropriate where $\boldsymbol{\eta}_0$ is a random vector with zero mean and unit variance and $\tau$ is the moment where steady state is established. Also, besides the diffusion, most of the quantum Monte Carlo methods implement branching of walkers, which reflects the influence of the potential term and is based on the estimate of the local energy which rules whether a given move is acceptable according to the Boltzmann distribution [5]. Although the branching greatly improves the convergence of the walk, we found that the TDQMC algorithm operates better without using it for more precise evaluation of the parameter $\alpha_j$ of Eq.(13) through minimization of the ground state energy. Once the value of $\alpha_j$ is determined the branching can be turned on in the standard way except that here we have birth and death of both particles and guide waves, based on the local energy given by:

$$E_{loc}^k = \sum_{i=1}^{N}\left[-\frac{\hbar^2}{2m}\frac{\nabla_i^2 \varphi_i^k(\mathbf{r}_i^k)}{\varphi_i^k(\mathbf{r}_i^k)} + V_{e-n}(\mathbf{r}_i^k)\right] + \sum_{i>j}^{N} V_{e-e}(\mathbf{r}_i^k - \mathbf{r}_j^k)\Bigg|_{\substack{\mathbf{r}_i^k = \mathbf{r}_i^k(t) \\ \mathbf{r}_j^k = \mathbf{r}_j^k(t)}} . \qquad (19)$$

In order to compare the TDQMC results with the numerically exact solution of the many-body Schrödinger equation here we employ a simplified model for one-dimensional helium atom with a soft core potential which has proven to be one of the most strongly correlated systems known (e.g. [16]; atomic units are used henceforth):

$$i\frac{\partial}{\partial t}\Psi(x_1,x_2,t)=\left[H_0+V_{e-e}\left(x_1-x_2\right)+V_{ext}\left(x_1,x_2,t\right)\right]\Psi(x_1,x_2,t), \tag{20}$$

where the "free" Hamiltonian reads:

$$H_0=-\frac{1}{2}\frac{\partial^2}{\partial x_1^2}-\frac{1}{2}\frac{\partial^2}{\partial x_2^2}-\frac{a}{\sqrt{1+x_1^2}}-\frac{a}{\sqrt{1+x_2^2}}, \tag{21}$$

and the electron-electron interaction potential reads:

$$V_{e-e}\left(x_1-x_2\right)=\frac{b}{\sqrt{1+(x_1-x_2)^2}} \quad . \tag{22}$$

The parameters $a$ and $b$ determine the strength of the electron-nuclear and electron-electron interaction, respectively. For helium in s-state (anti-parallel spins) we have $a$=2 and $b$=1 in Eqs.(21), (22). Then, the set of TDQMC equations (Eq.(9)-Eq.(11)) reads:

$$i\frac{\partial}{\partial t}\varphi_i^k(x_i,t)=\left[-\frac{1}{2}\nabla_i^2-\frac{a}{\sqrt{1+x_i^2}}+V_{e-e}^{eff}\left(x_i,t\right)+V_{ext}\left(x_i,t\right)\right]\varphi_i^k(x_i,t), \tag{23}$$

$i$=1,2; $k$=1,…,M, where the effective electron-electron potential is given by:

$$V_{e-e}^{eff}\left(x_i,t\right)=\sum_{j\neq i}^{2}\frac{1}{Z_j^k}\sum_{l=1}^{M}\frac{b}{\sqrt{1+\left[x_i-x_j^l(t)\right]^2}}\exp\left(-\frac{\left|x_j^l(t)-x_j^k(t)\right|^2}{2\sigma_j^k\left(x_j^k,t\right)^2}\right) \tag{24}$$

where:

$$Z_j^k = \sum_{l=1}^{M} \exp\left( -\frac{\left| x_j^l(t) - x_j^k(t) \right|^2}{2\sigma_j^k \left( x_j^k, t \right)^2} \right). \tag{25}$$

We start here with the calculation of the ground state of the atom by minimizing the total energy $E = \sum_k^M E_{loc}^k$ by varying the parameter $\alpha_j$; $j$=1,2..Since the two electrons are in s-state we have $\alpha_1 = \alpha_2 = \alpha$. The size of the grid for this calculation is 30 a.u. and the imaginary time step is 0.1 a.u. where 200 time steps are sufficient to reach a stable ground state. Figure 2 depicts the minimization curve which indicates that the optimal value of $\alpha$ is indeed close to unity, as expected for such a simple quantum system where the spatial uncertainty determines the range of quantum non-locality. Next, Fig. 3(a) shows a snapshot of the walker distribution at the end of the ground state calculation for $\alpha = 0.9$ while Fig. 3(b) shows that distribution smoothed by a kernel density estimation (red line) to be compared with the exact solution obtained from Eq.(20)-Eq.(22) (blue line). It is seen from Fig. 3(b) that the two distributions match quite satisfactory and that the TDQMC calculation correctly reflects the electron repulsion along the $x_1 = x_2$ axis (the main diagonal in Fig.3(b)). The total number of walkers and guide waves for this calculation is 64 800, distributed among 216 compute nodes where the ground state calculation takes less than a minute.

Next, in order to test the accuracy of TDQMC for ultrafast real-time evolution we carried out a calculation of the evolution of the one-body density matrix of the electron (see Eq.(15)) for real-time propagation where the nuclear potential $V_{e-n}$ is suddenly set to zero after the ground state preparation has ended (free diffraction of the electron waves, Figure 4):

$$\rho_1\left(x_1, x_1', t\right) = \frac{1}{M} \sum_{k=1}^{M} \varphi_1^{k*}(x_1, t) \varphi_1^k(x_1', t) \ , \tag{26}$$

where $\varphi_1^k \ (x_1, t)$ denotes the guide waves for the *k*-th walker. For comparison, we plot in Fig.4 also the numerically exact result (blue line) obtained from the direct integration of the two-body time-dependent Schrodinger equation (Eq.(20)) and next using the standard definition of the reduced (single-electron) density matrix expressed through the resulted wave-functions:

$$\rho\left(x,x',t\right)=\frac{\int \Psi(x,x_1,t)\Psi^*(x',x_1,t)dx_1}{\iint \Psi(x,x_1,t)\Psi^*(x,x_1,t)dx_1 dx} \tag{27}$$

It is seen from Fig.4(a,b) that the TDQMC result matches almost perfectly with the exact one which indicates that not only the distributions (Fig.3(b)) but also the coherences are calculated with good accuracy by the TDQMC algorithm, without referring explicitly to the many-body quantum state. In this calculation the space grid spans 50 a.u. and a total of 43 000 walkers and guide waves take part while the execution time is somewhat slower during the real-time propagation due to the calculation of two-dimensional quantities such as in Eq.(26), which are however not an essential part of the TDQMC algorithm.

The last numerical result here deals with the ionization and the degree of quantum coherence of one electron in one-dimensional helium atom exposed to a strong few cycle laser pulse, using the diagonal and the off-diagonal elements of the density matrix, Eq.(26), respectively. The external potential in dipole approximation reads $V_{ext}\left(x_i,t\right)=-x_i E\left(t\right)$ where $E\left(t\right)$ is the incoming pulse with amplitude 0.16 a.u., carrier frequency 0.1 a.u. and duration of two periods of the carrier. Physically, during each half cycle the electric field pushes the electron cloud away from the core which corresponds to the steps in the ionization curves in Fig.5(a), while at the same time the electron loses coherence due to its scattering from the core potential and the electron-electron interaction (Fig.5(b)). Here the degree of coherence is quantified as a sum of the elements along the main anti-diagonal of the density matrix. It is seen from Fig.5(a,b) that both the ionization and the degree of coherence stay in close correspondence with the exact result (blue lines) calculated from Eq.(20)-Eq.(22). This indicates that, despite the huge number of essentially stochastic wave functions involved in the TDQMC calculation, the elements of the resulting one-body density matrix of Eq. (15) are calculated with a very good accuracy.

## 4. Conclusions

It is shown here that concurrent ensembles of quantum trajectories in physical and in Hilbert spaces can be introduced within time-dependent quantum Monte Carlo method such that the predicted spatial distributions of walkers and one-body density matrices match well with the exact solution. The calculated density matrix correctly predicts the decoherence resulting from free diffraction of repulsing electrons and due to the ionization of an electron in a two electron atom by a powerful external field. This implies that by creating sufficiently rich statistics of walkers in physical space and in Hilbert space for each quantum species one can facilitate building an efficient approach to solving many-body time-dependent quantum problems. Our calculations reveal the although TDQMC is essentially a stochastic method it can perform well for both imaginary time and real time propagation as compared to other (imaginary time) methods using stochastic auxiliary fields. The stochasticity of the electron-electron potential is introduced through the random motion of the Monte Carlo walkers in TDQMC which makes it milder, as compared to in auxiliary field quantum Monte Carlo, which greatly improves the stability of the time propagation, allowing also real-time evolution. The TDQMC algorithm offers an affordable time scaling which is almost linear with the number of electrons, which is decisive for larger systems. This is owing to the intrinsic parallelism where the computational load is distributed almost independently between the compute nodes where the only communication needed is to calculate the nonlocal quantum correlations. Since the mass of the nucleus is much bigger than the electron mass, here we have considered the former to be a classical particle and thus we have neglected the electron-nuclear quantum correlations which, however, can be accounted for readily by including the nuclear wave functions in the product of Eq.(3), [17].

## 5. Acknowledgment

The author gratefully acknowledges support from the Bulgarian National Science Fund under grant FNI T02/10.

**Figure captions:**

**Figure 1**. Schematic of the transition from the mean-field approximation where all Monte Carlo walkers are guided by the same wave function to the TDQMC representation where each walker is guided by its own guide wave.

**Figure 2**. Energy of the ground state of 1D helium atom as function of the parameter $\alpha$ of Eq.(13), with the corresponding error bars (for 20 consecutive runs). The rightmost point shows the transition to the mean-field ground state energy.

**Figure 3.** Snapshot of the walker distribution for the ground state (a) and the corresponding smoothed distribution (b)-red line. The blue line in (b) depicts the exact ground state.

**Figure 4.** Contour maps of the moduli of the density matrices for the ground state (a) and after free diffraction of the interacting electron waves (b). Exact result – blue contours, TDQMC result – red contours.

**Figure 5**. Survival probability for a single electron in one-dimensional helium atom exposed to a powerful ultra-short laser pulse – (a), and the quantum coherence calculated as sum of the anti-diagonal elements of the density matrix in coordinate representation – (b). Red lines – from TDQMC, blue lines – exact result.

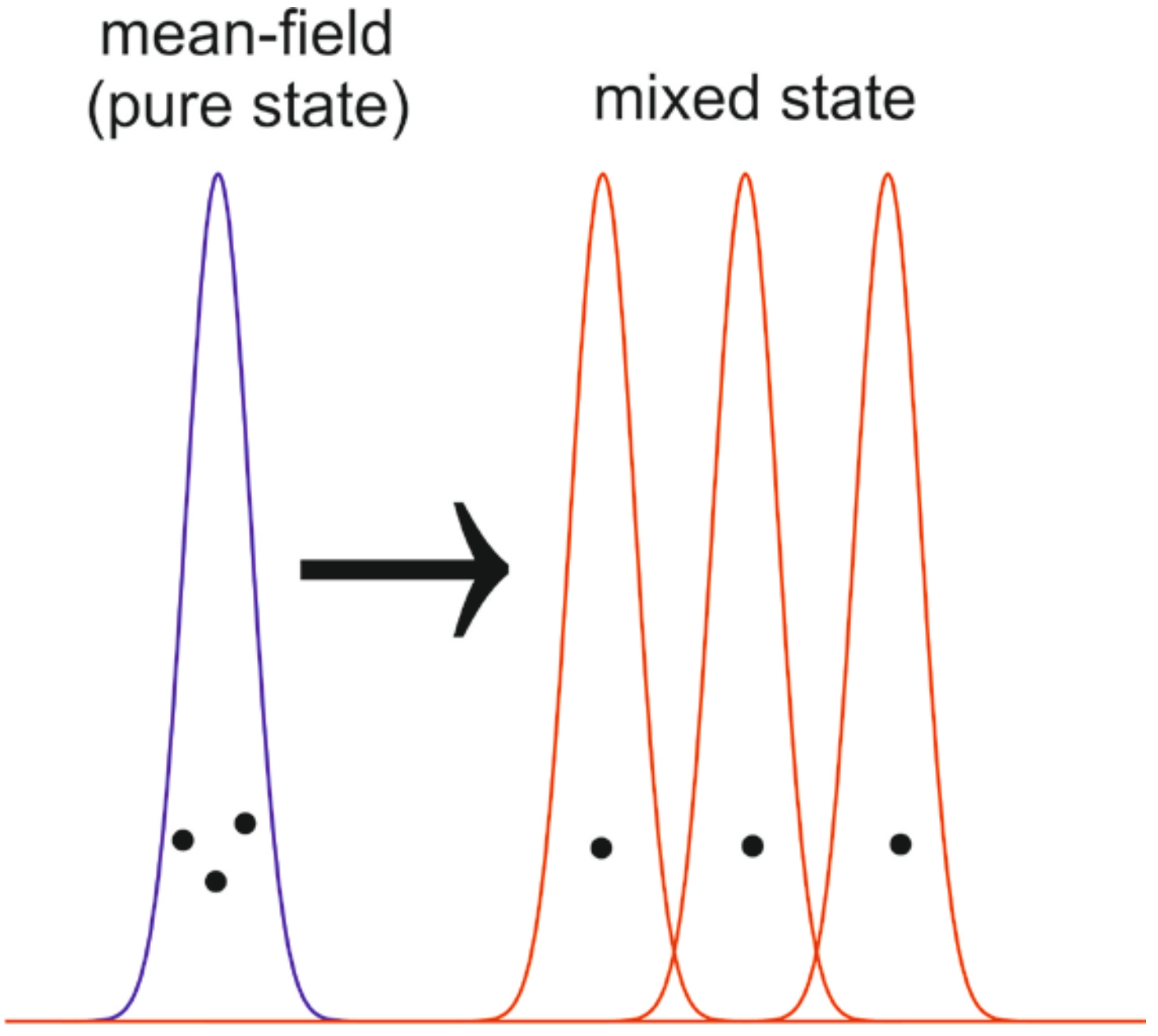

Christov, Figure 1

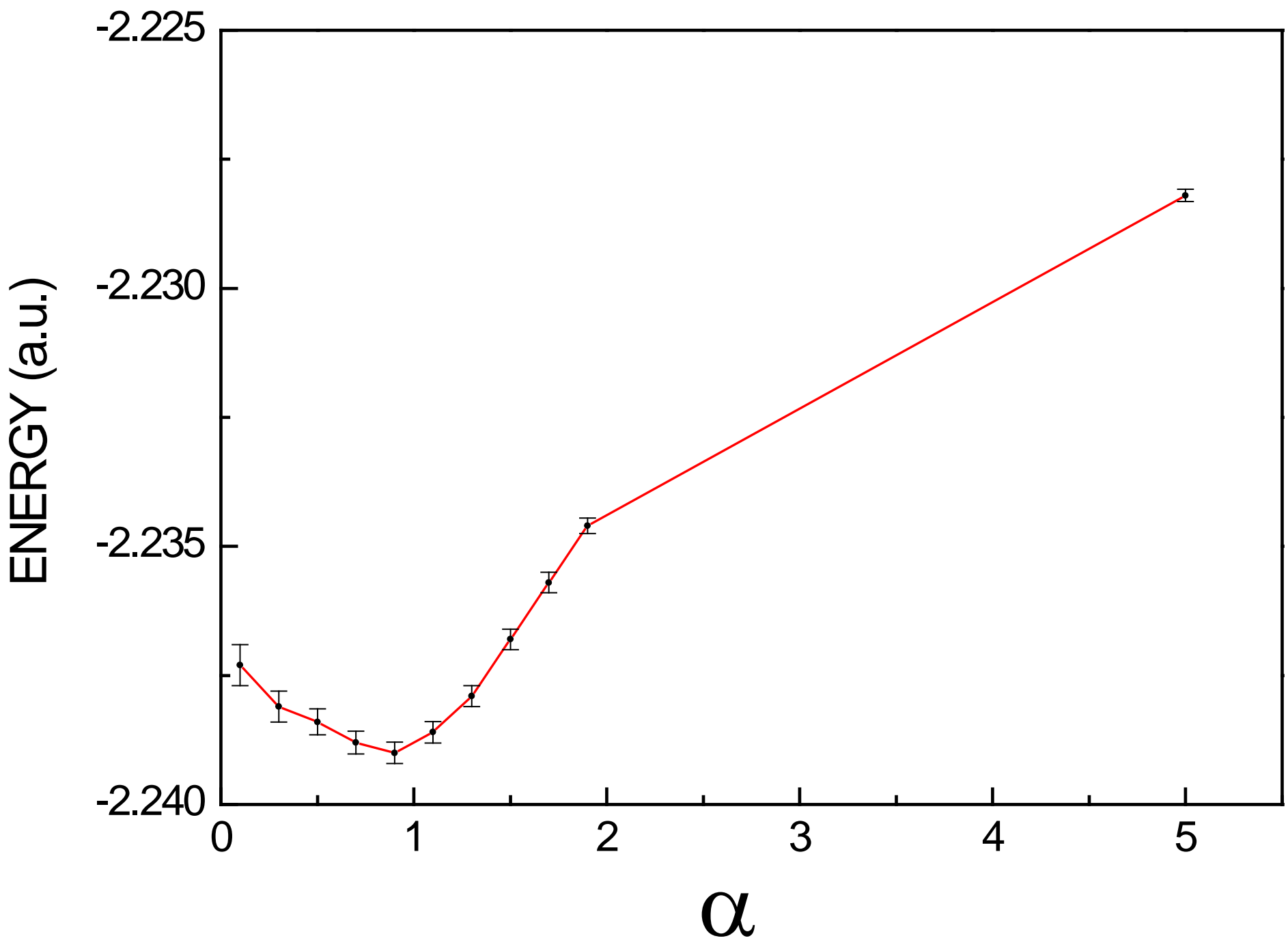

Christov, Figure 2

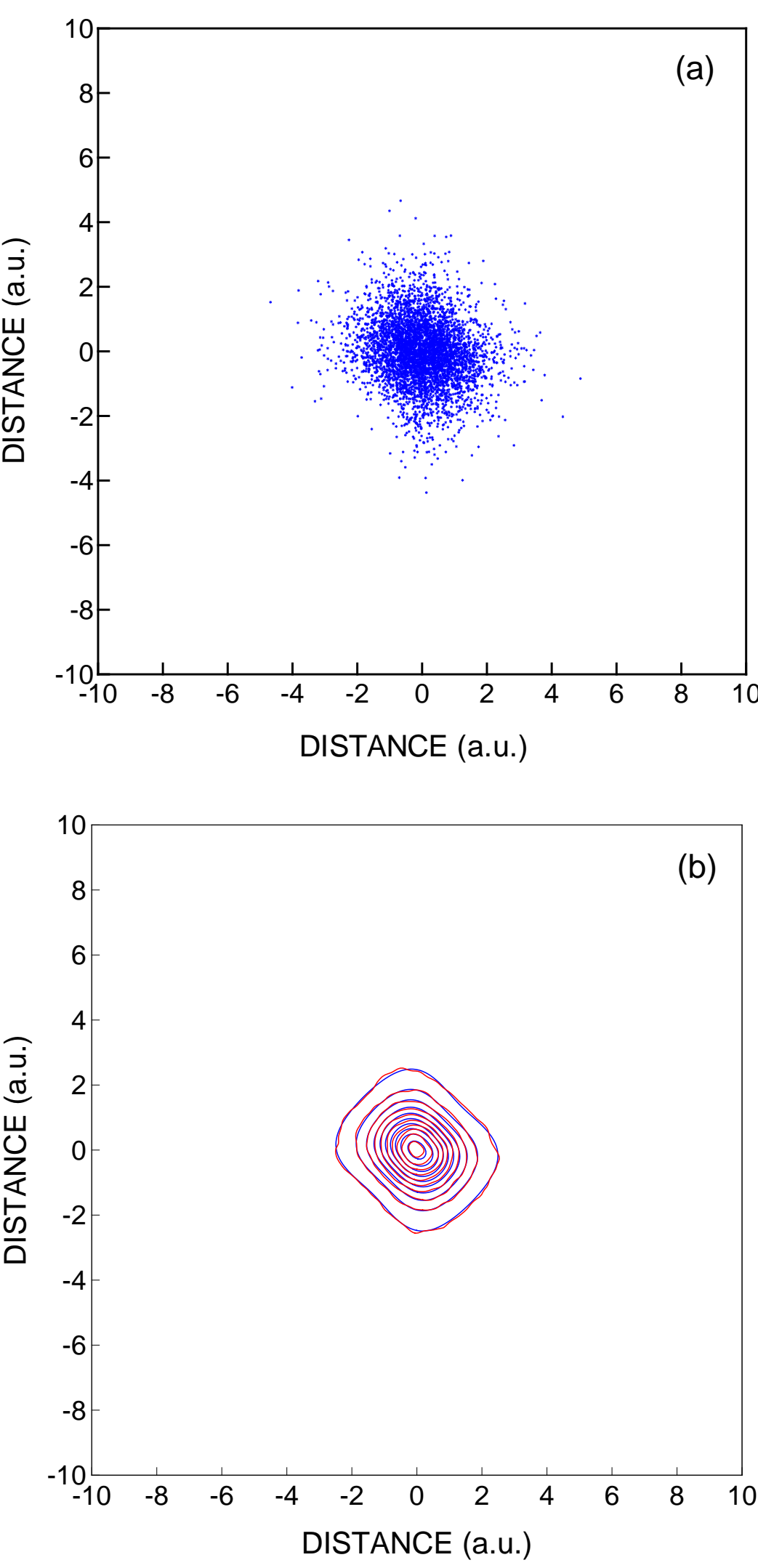

Christov, Figure 3

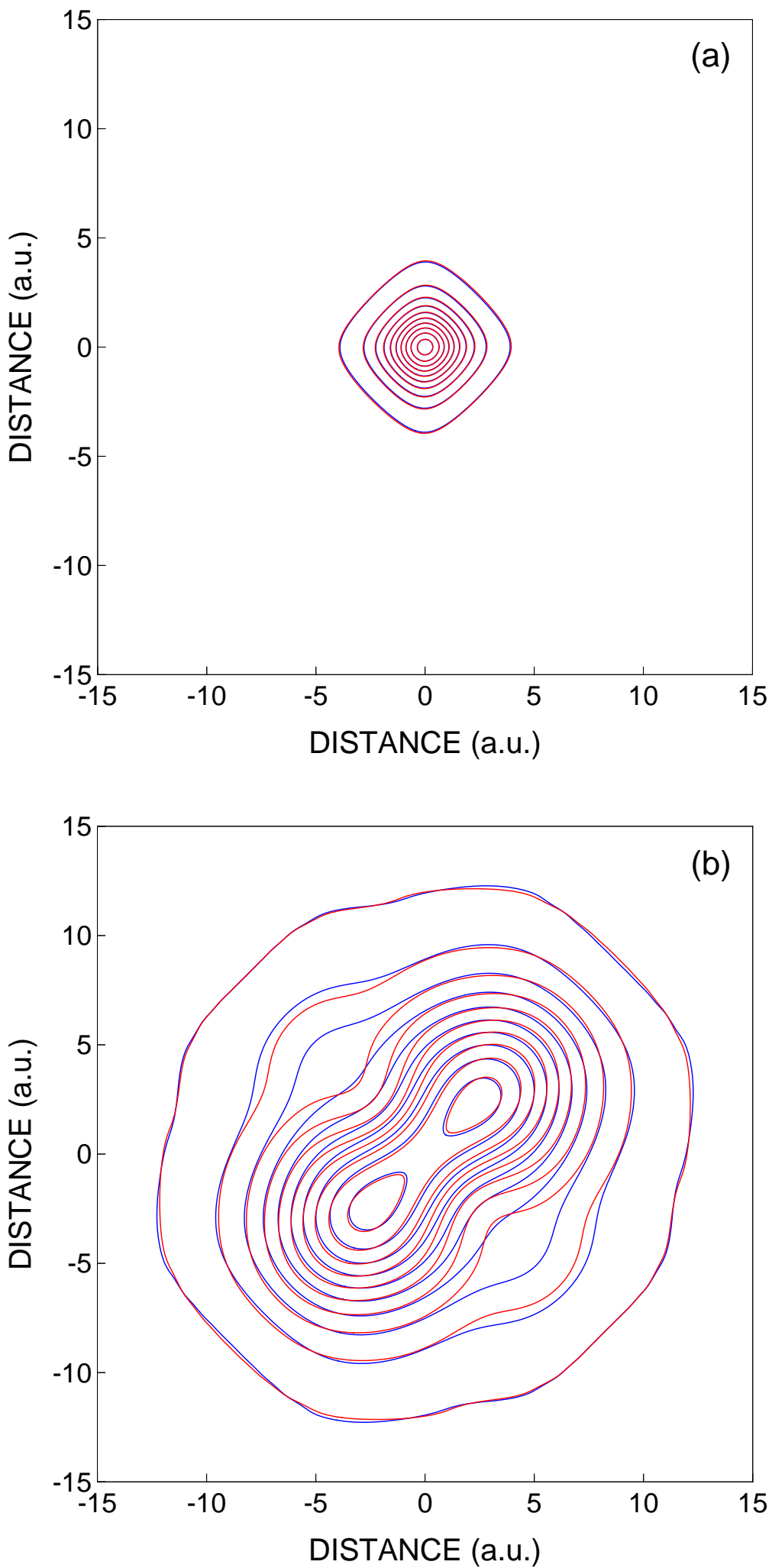

Christov, Figure 4

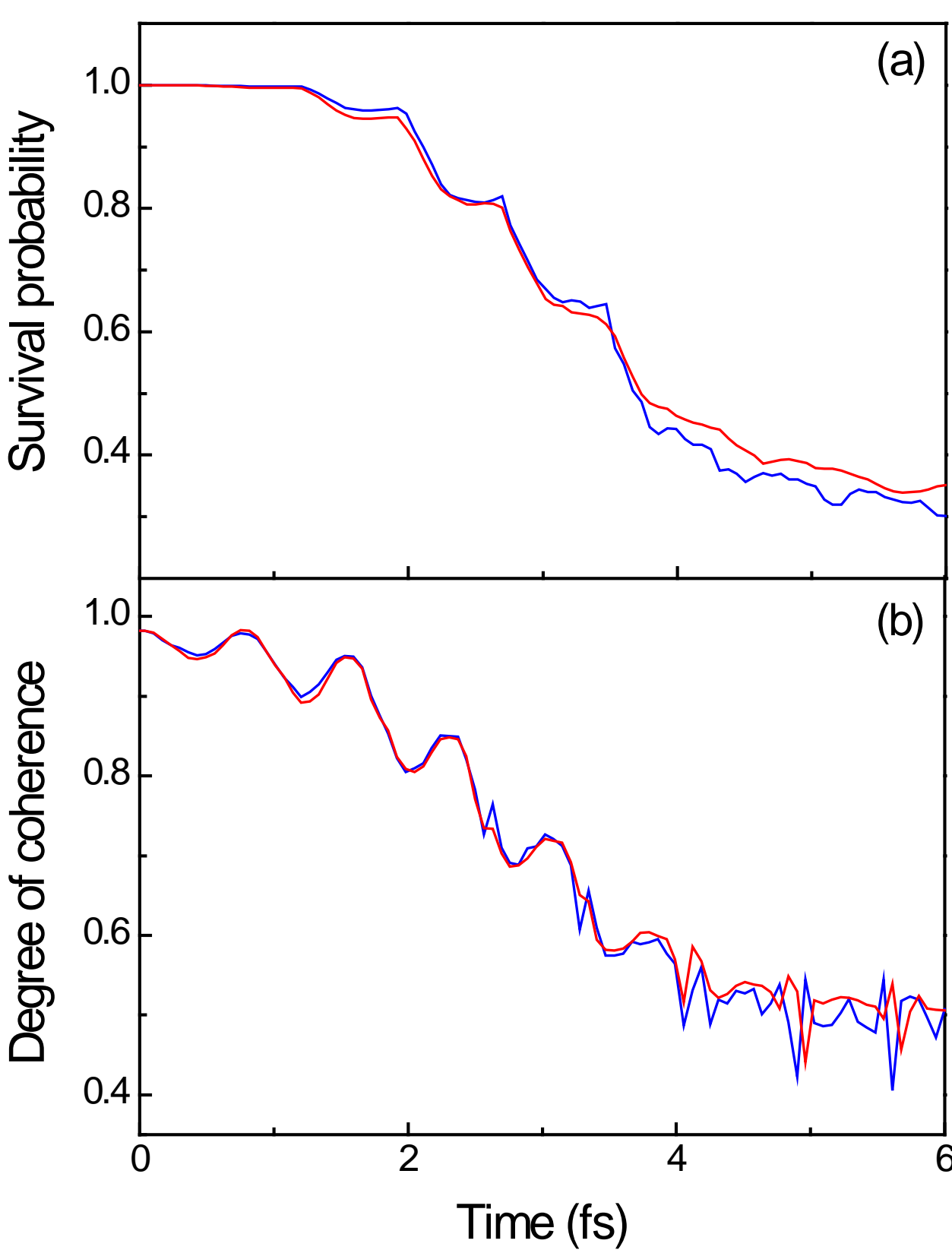

Christov, Figure 5